\documentclass[pre,aps,twocolumn,showpacs]{revtex4-2}
\usepackage{graphicx}% Include figure files
\usepackage{dcolumn}% Align table columns on decimal point
\usepackage{bm}% bold math
\usepackage{color}
\usepackage{amssymb}
\usepackage{amsmath}
\usepackage [latin1]{inputenc}
\usepackage{float}
\usepackage{tikz}

\newcommand\VRule[1][\arrayrulewidth]{\vrule width #1}

\newcommand{\tikzcircle}[2][red,fill=red]{\tikz[baseline=-0.5ex]\draw[#1,radius=#2] (0,0) circle ;}%

\begin{document}

\title{DNA Barcodes using a Double Nanopore System}

\author{Swarnadeep Seth}
\author{Aniket Bhattacharya}

\altaffiliation[]
{Author to whom the correspondence should be addressed}
{}
\email{Aniket.Bhattacharya@ucf.edu}
\affiliation{$^1$Department of Physics, University of Central Florida, Orlando, Florida 32816-2385, USA}
\date{\today}

\begin{abstract}
The potential of a double nanopore system to determine DNA barcodes has been  demonstrated experimentally. By carrying out Brownian dynamics simulation on a coarse-grained model DNA with protein tag (barcodes) at known locations along the chain backbone, we demonstrate that due to large variation of velocities of the chain segments between the tags, it is inevitable to under/overestimate the genetic lengths from the experimental current blockade and time of flight data. 
We demonstrate that it is the tension propagation along the chain's backbone that governs the motion of the entire chain and is the key element to explain the non uniformity and disparate velocities of the tags and DNA monomers under translocation that introduce errors in measurement of the length segments between protein tags.  Using simulation data we further demonstrate that it is important to consider the dynamics of the entire chain and suggest methods to accurately decipher barcodes. We introduce and validate an interpolation scheme using simulation data for a broad distribution of tag separations and suggest how to implement the scheme experimentally.
\end{abstract}

\maketitle

The use of digitized DNA-barcodes~\cite{barcode_CoxI,barcode_Hebert} in species identification~\cite{barcode_cryptic_species, barcode_taxonomy, barcode_seafood} has been a standard technique for the preservation of Earth's biological diversities~\cite{barcode_bio_diversity}.  The extinction of species is not homogeneous across the globe, rather a strong function of location. Many species in tropical countries are declining rapidly being on the verge of extinction.  The use of portable desktop equipment  will be beneficial to carry out the tests locally in different countries bypassing the restrictions of bringing samples from one country to a laboratory located in another country.  Nanopore based sequencing methods, such as, MinIon produced by Oxford nanopore~\cite{barcode_MiniON} is an important step towards that goal which will eventually replace traditional Sanger's type of sequencing. Thus there is a genuine need to develop real-time on site desktop methods for {\em in-situ} but fast and accurate determination of genetic information contained in barcodes. \par
\begin{figure}[ht!]
\includegraphics[width=0.45\textwidth]{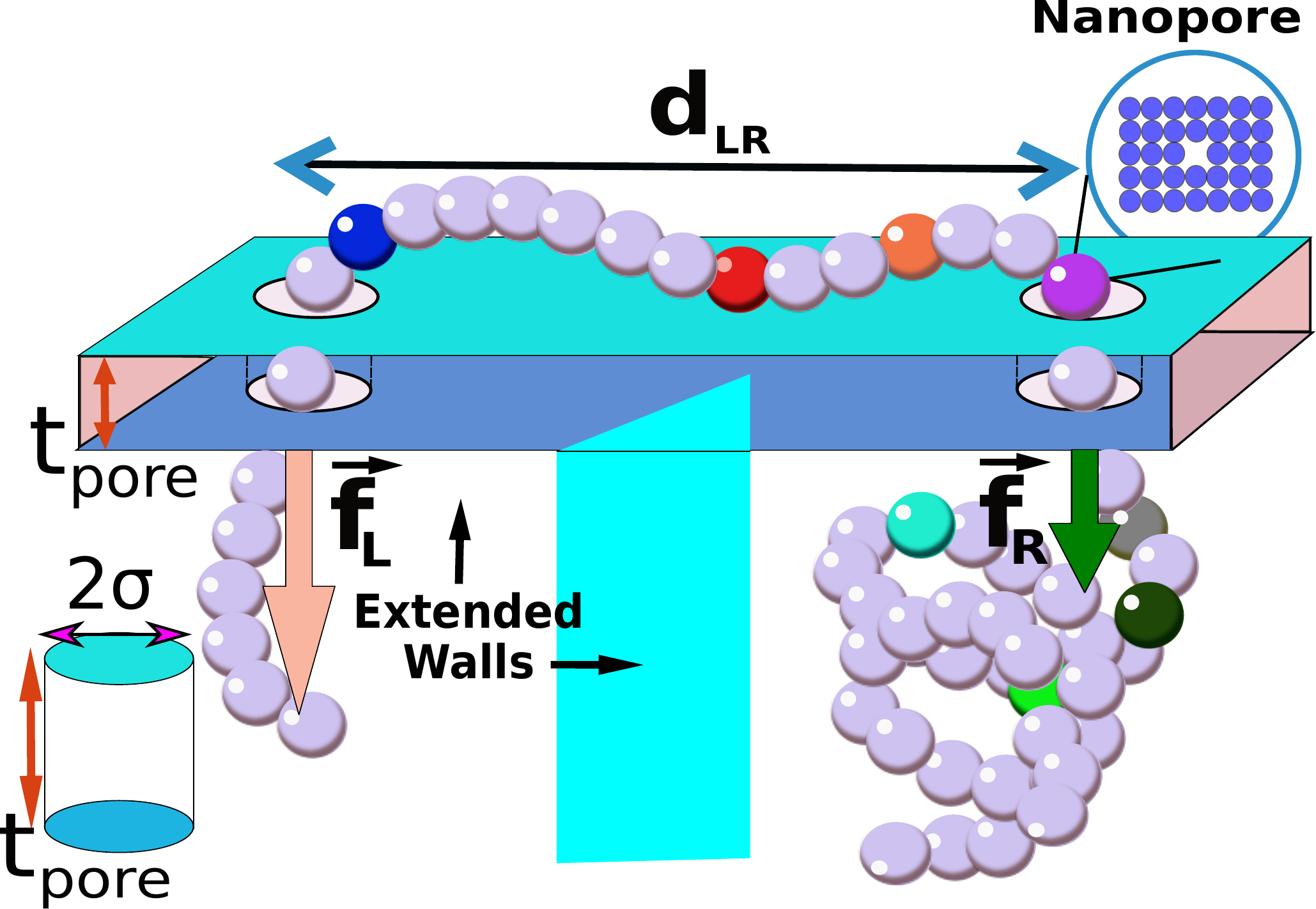}
\caption{\small \label{model} Schematic diagram of a dsDNA captured in between two nanopores drilled through an infinitely extended material with thickness $t_{pore}$ and at a separation distance $d_{LR}$. External bias forces $\vec{f}_L$ and $\vec{f}_R$  are applied to the monomers in the left and right nanopore respectively. The colored beads are tags attached to the nucleotides and have different mass and solvent friction different from the rest of the monomers. Keeping the force $\vec{f}_L$ to a fixed value, the force $\vec{f}_R$ is varied so that DNA in between the pore has a bias $\Delta \vec{f}_{LR} = \vec{f}_L - \vec{f}_R$ to floss it from one pore to the other.}
\end{figure}

A double nanopore platform (Fig.~\ref{model}) has demonstrated that it has the ability to outperform single nanopore based devices~\cite{Dekker-2016, Reisner-Small-2018}.  The captured DNA by both the pores is not only straightened by the tug-of-war forces present at each pore, but  adjustable differential biases and feedback mechanism at each pore offer overall a better control on the translocation speed~\cite{Reisner-Small-2019}. The most noteworthy aspect is the ability of multiple scans~\cite{Dekker-2016, Reisner-Small-2020} of the translocating DNA through the pores by flipping the net differential bias that not only increases the statistical accuracy of measurements, but in principle capable of providing additional information about the physical processes, which to date are largely unknown, and require a thorough theoretical investigation. \par
%%%%%%%%%%%%%%%%%%%%%%%%%%%%%%%%%%%%%%%%%%%%%%%%%%%%%%%%%%%%%
\begin{figure}[ht!]
\includegraphics[width=0.48\textwidth]{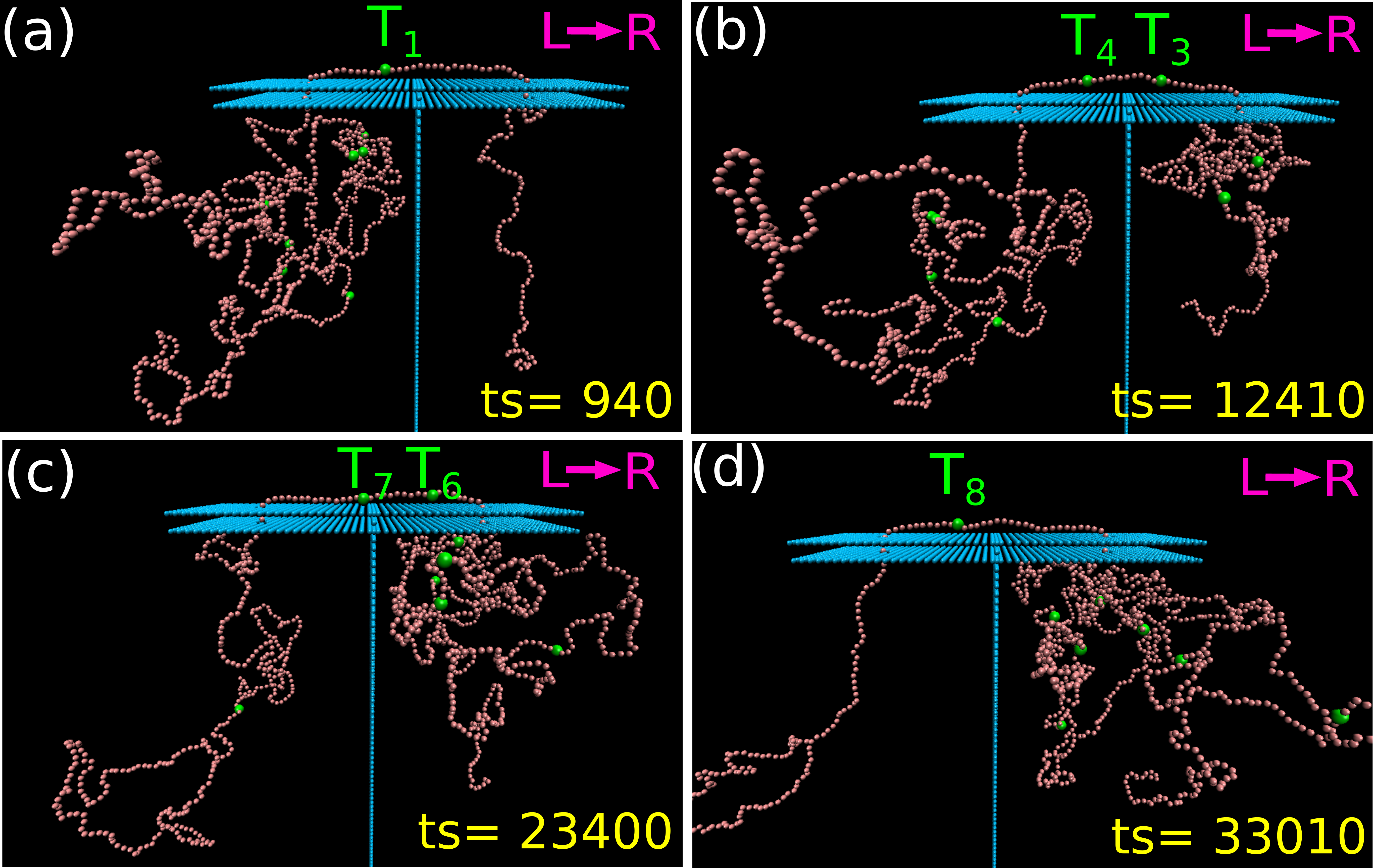}
\caption{\small\label{uncoiling}Figures (a)-(d) show the translocation of the tags from $L \rightarrow R$ direction during scanning. The pink beads \tikzcircle[fill=pink]{3pt} represent the normal monomers and the green blobs \tikzcircle[fill=green]{3pt} denote the heavier tags annexed to the chain. The progressive uncoiling of the dsDNA contour length on cis side and recoiling in the trans side which is a  consequence of the tension propagation (Fig.~\ref{tp}) seen from the simulation trajectory is important to be considered to calculate the barcodes accurately. Tags are depicted larger than the usual monomers for visualization purposes only. }
\end{figure}

While a double nanopore system offers immense promises, preliminary experiments reveal that extracting genomic lengths can be complicated due to lack of experimental information about the entire chain. The experiments can extract information about the dwell time of the protein tags in each pore and the time of flight (TOF) from one pore to the other only (please see Fig.~\ref{model}). These information would be sufficient to extract spacing between the barcode accurately if the entire chain was moving with the same velocity. As the protein tags are of different mass and experience different frictional drags, it is expected that different parts of the chain will not move through the double nanopore system with uniform velocity (please refer to Fig.~\ref{uncoiling}) which will then introduce error in genomic length determination.  The dispersion in velocity will also depend on the magnitude of the differential bias forces ($\Delta \vec{f}_{LR}$), the pore width ($t_{pore}$) and the distance between the two nanopores ($d_{LR}$). Experimentally nanoscale technology poses challenges to vary these parameters and can often be expensive. However, these dependencies can be efficiently studied in a model system using computer simulations
strategies~\cite{Bhattacharya_Seth_2020, Seth-JCP-2020, Aksimentiev-2020}  combined with theoretical methods developed in the last decade for studying translocating chain through
nanopores~\cite{Review1,Review2,Review3,Sakaue_PRE_2007} and validated by simulation studies~\cite{Ikonen_JCP2012, Adhikari_JCP_2013}
\par
%%%%%%%%%%%%%%%%%%%%%%%%%%%%%%%%%%%%%%%%%%%%%%%%%%%%%%%%%%%%%%%
%%%%%%%%%%%%%%%%%%%%%%%%%%%%%%%%%%%%%%%%%%%%%%%%%%%%%%%%%%%%%%%
In this letter, we present Brownian dynamics simulation results for a coarse-grained model of dsDNA with protein tags attached to it mimicking the essentials of the experimental setup. The simulation data shows that the velocity of the chain segments are indeed nonmonotonic. We further demonstrate that using the information about only the protein tags to extract barcode distances, as measured in an experiment, lead to an over/underestimation of the barcodes. We then use the nonequilibrium tension propagation theory of Sakaue~\cite{Sakaue_PRE_2007} to explain the non uniformity of the velocity profile. The underlying physical picture that emerges also provides clues for  the under/over estimation of the barcodes and direct us to an interpolation scheme to determine the barcodes accurately. 
\par
\begin{figure}[ht!]
\includegraphics[width=0.4\textwidth]{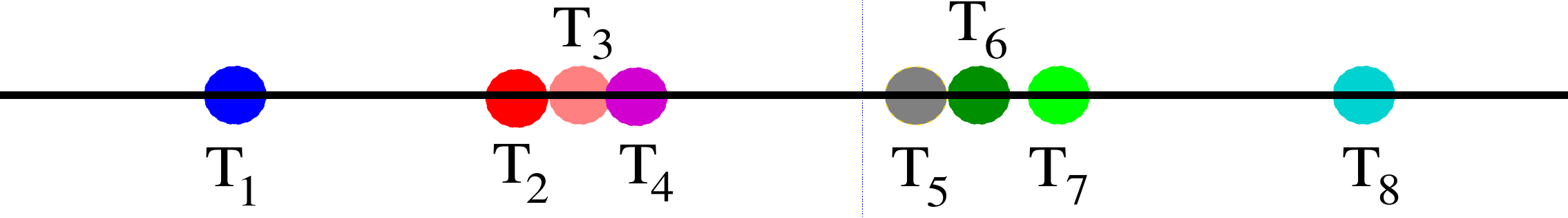}
\caption{\small \label{tags} Positions of the protein tags along the dsDNA.}
\end{figure}

$\bullet$~{\em Coarse-Grained Model and Brownian Dynamics}:~
Our coarse grained model consists of a polymer chain of 1024 beads with 8 protein tags translocating through a double nanopore system (Fig.~\ref{model}) inspired by the 48 kbp long double-stranded 
$\lambda$-DNA used in the experiment by  Liu {\em et al.}~\cite{Reisner-Small-2020}, where  sequence-specific protein tags are introduced chemically so that the distance between any two tags (Fig.~\ref{tags}) is known.
%#############################################################
% #############################################################
%%%%%%%%%%%%%%%%%%%%%%%%%%%%%%%%%%%%%%%%%%%%%%%%%%%%%%%%%%%%%%%
\begin{figure}[ht!]
\includegraphics[width=0.40\textwidth]{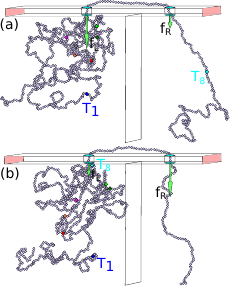}
\caption{\small \label{flossing}Demonstration of flossing the DNA keeping it captured in the double nanopore system. (a) The configuration shows a translocation taking place from $R \rightarrow L$ direction so that $\Delta\vec{f}_{LR} > 0$ and the last tag is on the right chamber. (b) Shows a snapshot a while later when the last tag has completed the translocation through both the pores, at which point the differential voltage is reversed so that $\Delta\vec{f}_{RL}>0$ and the translocation proceeds in the $L \rightarrow R$ direction.}
\end{figure}
%%%%%%%%%%%%%%%%%%%%%%%%%%%%%%%%%%%%%%%%%%%%%%%%%%%%%%%%%%%%%%%
\par
In simulation, each bead (monomer) represents approximately 46 bp unit long dsDNA and one tag is roughly equivalent to 75 bp in the experiment which translates to approximately 2 - 3 beads. The general scheme of the BD simulation strategy for a homo-polymer translocating through a double nanopore has been discussed in our recent publication~\cite{Bhattacharya_Seth_2020,Seth-JCP-2020}. The protein tags are introduced by choosing the mass and friction coefficient at tag locations to be significantly different from the rest of the monomers along the chain which requires modification of the BD algorithm~\cite{CNP-Barcode-2021}. Instead of explicitly putting side-chains at the tag locations, we made the mass and the friction coefficient of the tags 3 times larger. This we find enough to resolve the distances among the tags. The location of the tags are also chosen in such a way so that genetic distances are disparate (some tags being close by and some are far apart) as shown in Fig.~\ref{tags} and Table-I. It is also noteworthy that there is no left to right symmetry so that the center of mass of the chain is not located at the center of the chain. The key question to answer is mimicking  the double nanopore experimental protocol how accurately can one extract these genomic distances so that the method then can be applied to determine genetic lengths in  unknown specimens.
%###############################################################################################
\begin{table}[ht!]
\caption{Tag positions along the dsDNA}
\begin{tabular}{l l l l l l l l l}\hline  
Tag \# & \color{blue} \bf{}$T_1$ ~~& \color{red} \bf{}$T_2$ ~~& \color{orange} \bf{}$T_3$ ~~& \color{magenta} \bf{}$T_4$ ~~& \color{darkgray} \bf{}$T_5$ ~~& \color{teal} \bf{}$T_6$ ~~& \color{green} \bf{}$T_7$ ~~& \color{cyan} \bf{}$T_8$\\ \hline  
Position & 154 & 369 & 379 & 399 & 614 & 625 & 696 & 901 \\ \hline  
Separation &  154 & 215 & 10 & 20 & 215 & 11 & 71 & 205 \\ \hline 
\end{tabular}
\end{table}
\par
\begin{figure}[ht!]
\includegraphics[width=0.45\textwidth]{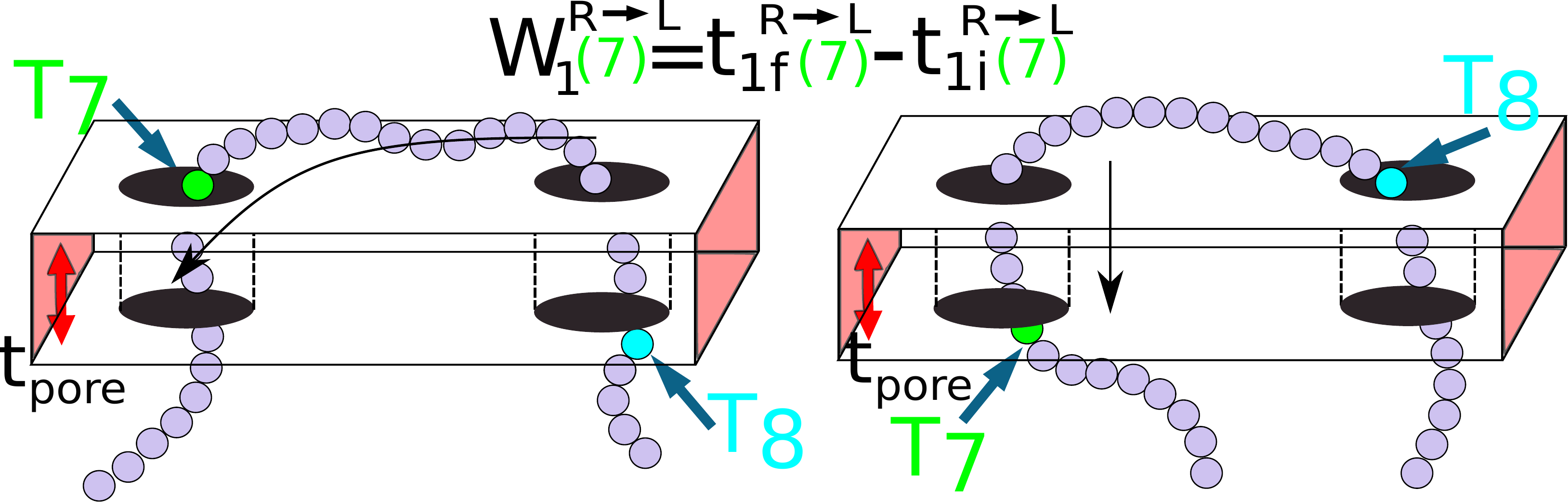}
\caption{\small \label{dwell-time} Dwell time of ${\bf\color{green} T_7}$ is calculated by recording the time  difference between arrival $t_{i}^{R \rightarrow L} (7)$ and exiting time $t_{f}^{R \rightarrow L} (7)$ within the left nanopore thickness for $R \rightarrow L$ translocation.}
\end{figure}
\begin{figure}[ht!]
\includegraphics[width=0.45\textwidth]{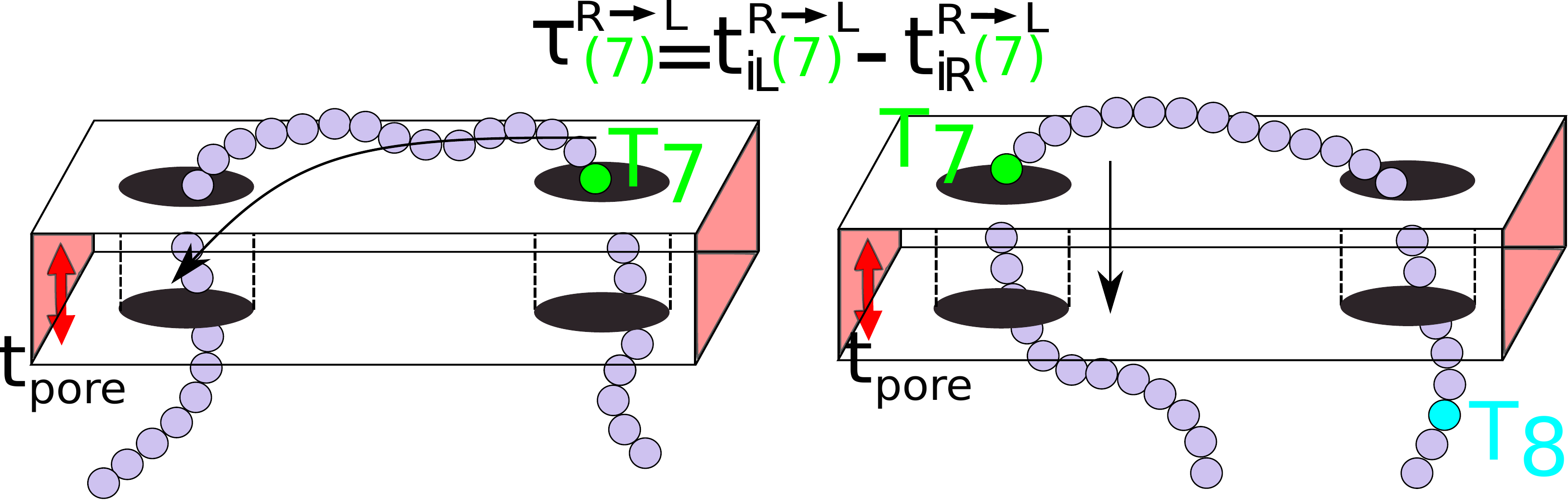}
\caption{\small \label{tof} Illustration depicts the time of flight ($tof$) of ${\bf\color{green} T_7}$ is measured as the time taken to reach to left pore from right pore for $R \rightarrow L$ motion.}
\end{figure}

$\bullet$~{\em Repeated scans and measurements}:~The measurement protocols are described in Figs.~\ref{flossing}, \ref{dwell-time}, and \ref{tof}. The differential bias  $\Delta\vec{f}_{LR}=\vec{f}_{L}-\vec{f}_R > 0$ for the $R \rightarrow L$ translocation (Fig.~\ref{flossing}(a)). Once the last tag passes through the right pore, the bias is switched to $\Delta\vec{f}_{RL}=\vec{f}_{R}-\vec{f}_L > 0$ (Fig.~\ref{flossing}(b)) so that the direction of translocation is reversed.  Here, we report the case for $|\Delta\vec{f}_{LR}|=|\Delta\vec{f}_{RL}|$. Later we mention what happens when an asymmetry is present and  $|\Delta f_{LR}| \ne |\Delta f_{RL}|$. This process (called flossing - one flossing consists of one $R \rightarrow L$ and one $L \rightarrow R$ scan) is repeated for 300 times and we record the experimentally measurable  quantities, - the  dwell time and the time of flight ($tof$)  as described next. We reserve the subscripts 1 and 2 for the left pore (pore 1) and right pore (pore 2) respectively. For the sake of brevity we define quantities for the $L \rightarrow R$ only, implicating that DNA is translocation from left pore to right pore and the same definitions hold for the $R \rightarrow L$ translocation. \par
$\bullet$~{\em Dwell time}:~
The co-captured dsDNA in a dual nanopore system provides two different ways of time measurements for the protein tags during translocation which can be translated to genomic lengths. Similar to a single nanopore setup, one can measure the
{\em dwell time}  (Fig~\ref{dwell-time}) calculated by recording the time difference between the arrival $t_i^{R \rightarrow L}(m)$ and the exit time $t_f^{R \rightarrow L}(m)$ of a monomer $m$, defined for the $R \rightarrow L$ translocation as follows:
\begin{subequations}
\label{dwell}
\begin{gather}
W_1^{L \rightarrow R}(m) = t_{1f}^{L \rightarrow R}(m) - t_{1i}^{L \rightarrow R}(m) \\
W_1^{R \rightarrow L}(m) = t_{1f}^{R \rightarrow L}(m) - t_{1i}^{R \rightarrow L}(m)
\end{gather}
\end{subequations}
Likewise, $W_2^{L \rightarrow R}(m)$ and $W_2^{R \rightarrow L}(m)$ can be obtained replacing $1$ by $2$ in the above equation. An example of dwell time calculation for ${\bf\color{green} T_7}$  is shown in Fig.~\ref{dwell-time}. 
%%%%%%%%%%%%%%%%%%%%%%%%%%%%%%%%%%%%%%%%%%%%%%%%%%%%%%
%%%%%%%%%%%%%%%%%%%%%%%%%%%%%%%%%%%%%%%%%%%%%%%%%%%%%%%%%%%%%%%
The dwell time translates to experimental {\it current blockade} of monomers. \par
$\bullet$~{\em The time of flight (tof)}:~
In addition to the dwell time measurements, in a double nanopore setup, one can also measure the  time taken by a monomer as it leaves one pore and touches the upper boundary of the other pore during its voyage across the pore separation $d_{LR}$ (Fig.~\ref{tof}). This is called the time of flight $tof$ and defined as follows:
\begin{subequations}
  \label{tof-eqn}
\begin{gather}
\tau^{L \rightarrow R}(m) = t_{iR}^{L \rightarrow R}(m) - t_{iL}^{L \rightarrow R}(m) \\
\tau^{R \rightarrow L}(m) = t_{iL}^{R \rightarrow L}(m) - t_{iR}^{R \rightarrow L}(m).
\end{gather}
\end{subequations}
%%%%%%%%%%%%%%%%%%%%%%%%%%%%%%%%%%%%%%%%%%%%%%%%%%%%%%%%%%%%%%%
\par
%%%%%%%%%%%%%%%%%%%%%%%%%%%%%%%%%%%%%%%%%%%%%%%%%%%%%%%%%%%%%%% 
$\bullet$~{\em Dwell velocity $v_D$ \& time of flight velocity $v_{tof}$}:~Accordingly, one can calculate both $v_D$ as well as $v_{tof}$ using Eqns.~\ref{dwell-time} and \ref{tof} as follows,
\begin{subequations}
  \begin{gather}
    \label{v_D1}
v_{D}^{L \rightarrow R}(m) = \frac{1}{2}\left[ \frac{t_{pore}}{W_1^{L \rightarrow R}(m)}  +   \frac{t_{pore}}{W_2^{L \rightarrow R}(m)} \right]
\\
  \label{v_D2}
  v_{D}^{R \rightarrow L}(m) = \frac{1}{2}\left[ \frac{t_{pore}}{W_1^{R \rightarrow L}(m)}  +   \frac{t_{pore}}{W_2^{R \rightarrow L}(m)} \right]\\
    \label{v_D12}
\langle v_D(m) \rangle = \frac{1}{2}\left ( \langle v_{D}^{L \rightarrow R}(m) \rangle + \langle v_{D}^{R \rightarrow L}(m) \rangle \right)
\end{gather}
\end{subequations}
%###############################################################
\begin{subequations}
\begin{gather}
v_{tof}^{L \rightarrow R}(m) = d_{LR}/\tau^{L \rightarrow R}(m) 
\\
v_{tof}^{R \rightarrow L}(m) = d_{LR}/\tau^{R \rightarrow L}(m)\\
\langle v_{tof}(m) \rangle = \frac{1}{2}\left (  \langle v_{tof}^{L \rightarrow R}(m) \rangle  +  \langle v_{tof}^{R \rightarrow L}(m) \rangle \right)
\end{gather}
\end{subequations}
Here $\langle \cdot\cdot\cdot \rangle$ implies average over multiple scans through the left and the right pore which reduces the statistical error in the measurements.
\par
\begin{figure}[h!]
\includegraphics[width=0.45\textwidth]{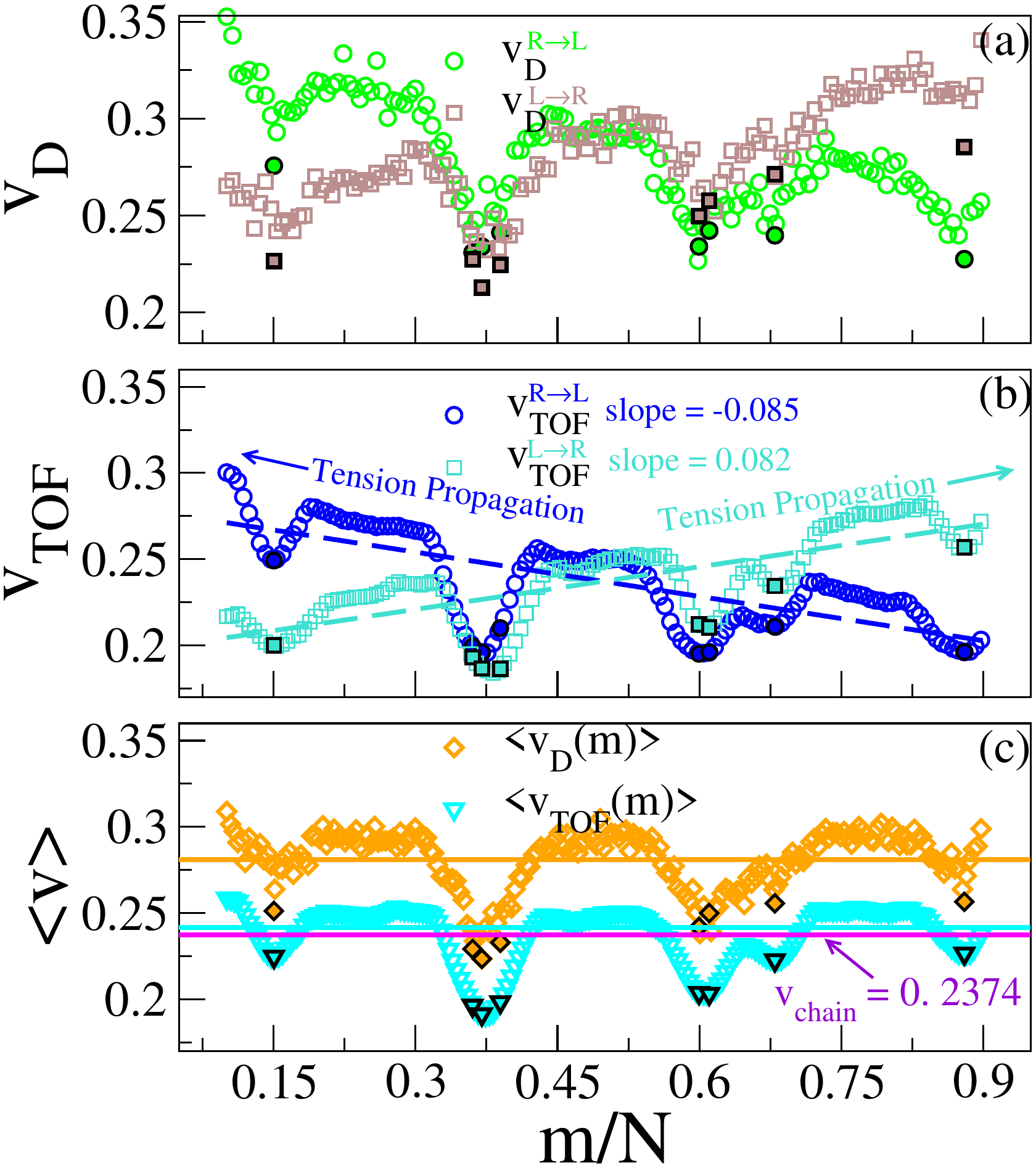}
\caption{\small \label{velocity} (a) Dwell velocities and (b) TOF velocities of the monomers during $R \rightarrow L$ ({\bf\color{blue} \large $\circ$}) and $L \rightarrow R$ ({\bf\color{cyan} \tiny $\square$}) translocation. Eight tag velocities are marked by the filled symbols. The dotted lines (Turquoise and Blue) indicate the directions of the tension propagation for $ L \rightarrow R$ and  $ R \rightarrow L$ respectively.  (c) Directional averaged dwell velocity and TOF velocities  are represented with {\bf\color{orange} \large $\diamond$} and {\bf\color{cyan} $\triangledown$}. The solid {\bf\color{orange} orange} horizontal line represents average dwell velocity and the {\bf\color{cyan} cyan} line represents the average tof velocity of all monomers respectively. The average velocity of the entire chain is represented by the {\bf\color{violet} violet} line. }  
\end{figure}

{\em $\bullet$~Non uniform velocity profile:} Due to lack of $L \rightarrow R$ symmetry both $v_{D}$ and $v_{tof}$ has a positive slope along the direction of translocation due to propagation of the tension front which can be explained using tension propagation theory~\cite{Ikonen_JCP2012} (Fig.~\ref{velocity}(a) and (b)). Fig.~\ref{velocity}(c) shows that average of  both the direction and the slope is close to zero as we are considering the symmetric differential bias  $|\Delta\vec{f}_{LR}|=|\Delta\vec{f}_{RL}|$. As expected, the  velocity distribution of the chain segments become non-monotonic. The protein tags with heavier mass ($m_{tag} > m_{bulk}$) and larger solvent friction ($\gamma_{tag} > \gamma_{bulk}$) reside at the lower envelope of the graphs (Fig.~\ref{velocity}(c)) while the dsDNA monomers reach their maximum velocity somewhere in between the  tags. In addition to the average dwell and time of flight velocities $\langle v_{D}\rangle$ and  $\langle v_{tof}\rangle$, we have plotted the average velocity of the entire chain $\langle v_{chain}\rangle$, each represented as a solid line. For the choice of the parameters the dwell time velocities are larger and $\langle v_{D}\rangle $ is about 20\% larger than the $\langle v_{tof}\rangle$. This is a coincidence and not a generic feature. We have checked that keeping all the parameters the same, an increase in pore thickness $t_{pore}$ will result in an overall decrease of the dwell time velocities due to increased friction for a thicker pore. It is also worthwhile to note that although in simulation we can calculate the dwell time  and time of flight for all the monomers and tags, experimentally these data (Eqns.~\ref{dwell} and \ref{tof-eqn}) are measured for the protein tags only as the tags produce significant current blockades to be measured. However, the entire chain contributes to the dynamics of the tags, and it is this lack of information for the entire chain inevitably leads to inaccurate measurements of the genomic length as demonstrated below. 
\par
\begin{figure}[ht!]
\includegraphics[width=0.45\textwidth]{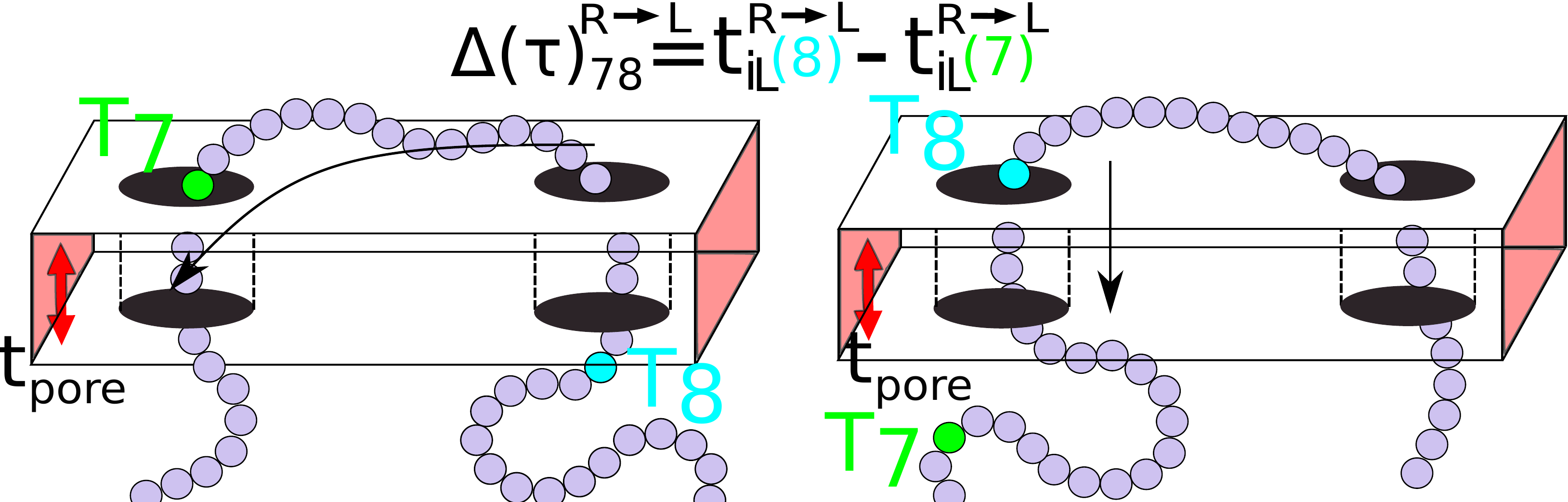}
\caption{\small \label{time-delay} Demonstration of calculation of tag time delay $\Delta(\tau)_{78}^{R \rightarrow L}= t_{iL}^{R \rightarrow L}(8) - t_{iL}^{R \rightarrow L}(7)$ for tags {\bf\color{green} $T_7$}  and {\bf\color{cyan} $T_8$} at the left pore. The similar quantity for $L \rightarrow R$ translocation  $\Delta(\tau)_{87}^{L \rightarrow R}= t_{iR}^{L \rightarrow R}(7) - t_{iR}^{L \rightarrow R}(8)   \ne \Delta(\tau)_{78}^{R \rightarrow L} $ as of the asymmetric tag positions along the chain.}
\end{figure}
\begin{figure}[ht!]
\centering 
\includegraphics[width=0.4\textwidth]{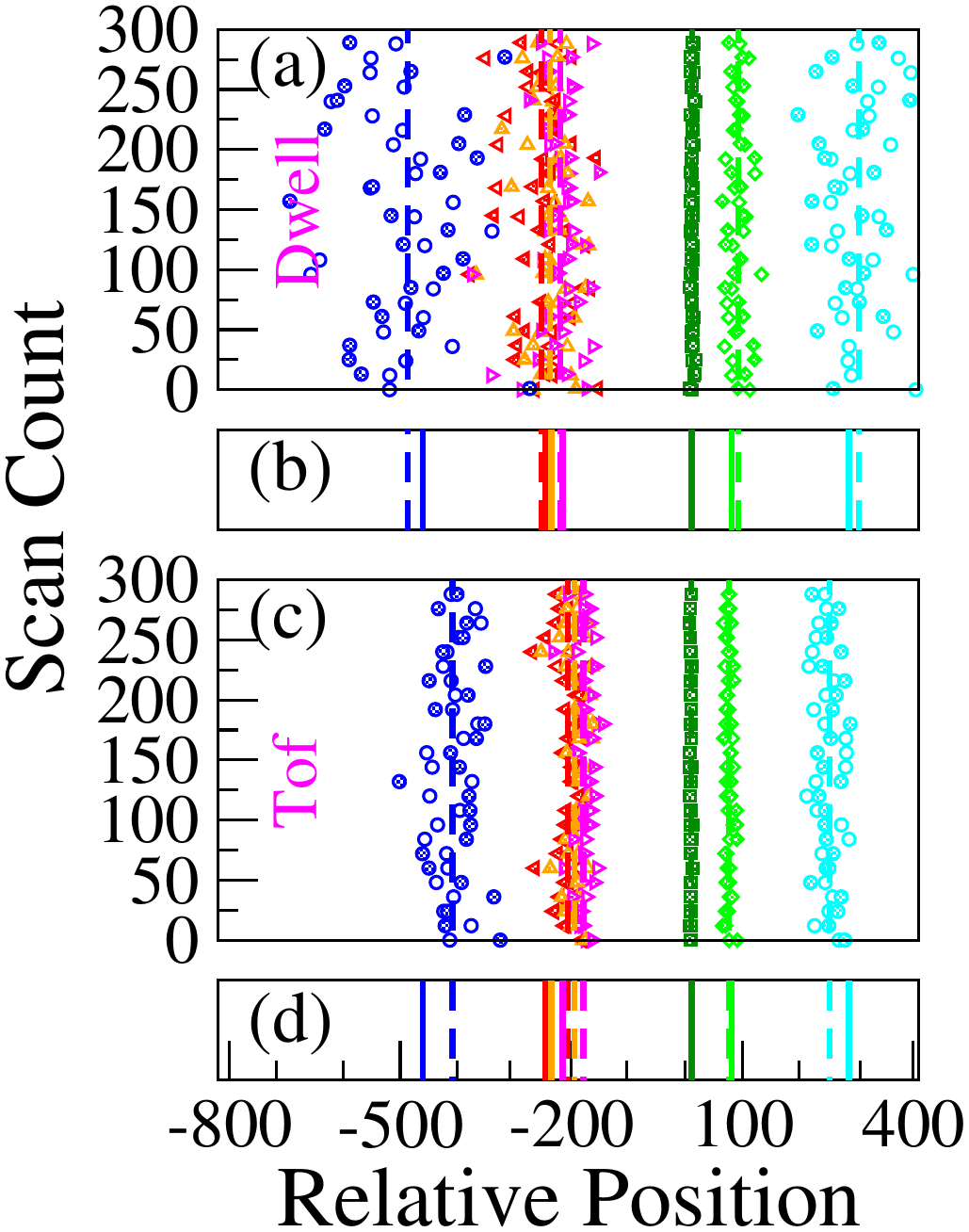}
\caption{\small \label{BC-dwell} Barcode generated using (a) Eqn.~\ref{d_dwell} and (c) Eqn.~\ref{dmn3}. Colored symbols/lines refer to the barcodes {\bf \color{blue} $T_1$}, {\bf \color{red} $T_2$ }, {\bf \color{orange} $T_3$}, {\bf \color{magenta}, $T_4$}, {\bf \color{teal} $T_6$}, {\bf \color{green} $T_7$}, and {\bf \color{cyan} $T_8$} calculated w.r.t {\bf \color{darkgray} $T_5$}. The open/filled symbols correspond to $R/L\rightarrow L/R$ translocation. For better visualization, every 6th data point is shown. The solid/dashed colored lines in (b) and (d) are the exact and the calculated barcodes averaged over $300$ scans.}
\end{figure}

%\vskip -0.5truecm
%%%%%%%%%%%%%%%%%%%%%%%%%%%%%%%%%%%%%%%%%%%%%%%%%%%%%%%%%%%%
$\bullet$~{\em Barcodes from the segmental velocity connecting two tags:~}Let us first evaluate the consequence if we calculate the average velocity of the segment $v_{seg}^{L \rightarrow R} (m,n)$ connecting the tags $T_m$ and $T_n$ by  approximating it to be the average velocity of the tags  $T_m$ and $T_n$ only so that 
\begin{subequations}
\begin{gather}
\label{vmn}
v_{seg}^{L \rightarrow R} (m,n) \approx \frac{1}{2} \left[v_{tof}^{L \rightarrow R}(m) + v_{tof}^{L \rightarrow R}(n)\right]\\
v_{seg}^{R \rightarrow L} (m,n) \approx \frac{1}{2} \left[v_{tof}^{R \rightarrow L}(m) + v_{tof}^{R \rightarrow L}(n)\right]
\end{gather}
\end{subequations}
Barcode distances $d_{tof}^{L \rightarrow R} (m,n)$ and $d_{tof}^{R \rightarrow L} (m,n)$  between $T_m$ and $T_n$ for $L(R) \rightarrow R(L)$ translocations are then obtained as
\begin{subequations}
\begin{align}
\label{dmn1}  
d_{tof}^{L \rightarrow R} (m,n) & \approx  v_{mn}^{L \rightarrow R} (tof) \times (\Delta \tau)_{mn}^{L \rightarrow R},\\
\label{dmn2}  
d_{tof}^{R \rightarrow L} (m,n) & \approx  v_{mn}^{R \rightarrow L} (tof) \times( \Delta \tau)_{mn}^{R \rightarrow L},\\
\label{dmn3}  
d_{tof}(m,n) & \approx  \frac{1}{2}\left( d_{tof}^{L \rightarrow R} (m,n) +  d_{tof}^{R \rightarrow L} (m,n) \right).
\end{align}
\end{subequations}
Here, $(\Delta \tau)_{mn}^{L \rightarrow R}$ and $(\Delta \tau)_{mn}^{R \rightarrow L}$ are the time difference  of arrival (we call it {\em tag-time-delay}) of the $m$-th and the $n$-th tags at $L/R$ pore during $R/L \rightarrow L/R$ translocation. Fig.~\ref{time-delay} illustrates
a specific case $(\Delta\tau)_{78}^{R \rightarrow L}$.
%%%%%%%%%%%%%%%%%%%%%%%%%%%%%%%%%%%%%%%%%%%%%%%%%%%%%%%%%%%%%%%
%%%%%%%%%%%%%%%%%%%%%%%%%%%%%%%%%%%%%%%%%%%%%%%%%%%%%%%%%%%%%%%
Eqn.~\ref{dmn3} provides the final distance $d_{tof}(m,n)$ averaged over multiple scans in each direction. 
Likewise, using dwell time velocity Eqns.~\ref{v_D1}-\ref{v_D12}, and replacing the subscript {\em tof} by {\em dwell} in Eqn~\ref{vmn}, one can derive equations analogous to Eqns.~\ref{dmn1}-\ref{dmn3}. A similar equation for the barcode distance using dwell time data from both the pores: 
\begin{equation}
\label{d_dwell} 
d_{dwell}(m,n)  \approx  \frac{1}{2}\left( d_{dwell}^{L \rightarrow R} (m,n) +  d_{dwell}^{R \rightarrow L} (m,n) \right)  
\end{equation}
The distribution of barcode distances with respect to ${\bf\color{gray} T_5}$ using Eqns.~\ref{dmn3} and \ref{d_dwell} are shown in Fig.~\ref{BC-dwell} and summarized in the $4^{th}$ and $3^{rd}$ columns of Table-II respectively. 
%%%%%%%%%%%%%%%%%%%%%%%%%%%%%%%%%%%%%%%%%%%%%%%%%%%%%%%%%%%%%%%
%%%%%%%%%%%%%%%%%%%%%%%%%%%%%%%%%%%%%%%%%%%%%%%%%%%%%%%%%%%%%%%
\begin{table*}[ht!]
\centering
\caption{Velocity to Barcodes using different Methods}
\begin{tabular}{
	!{\color{violet}\VRule[1pt]}c 
	!{\color{violet}\VRule[1pt]}c 
	!{\color{violet}\VRule[1pt]}c 
	!{\color{red}\VRule[1pt]}c
	!{\color{green}\VRule[1pt]}c  
	!{\color{green}\VRule[1pt]}c 
	!{\color{green}\VRule[1pt]}c 
	!{\color{cyan} \VRule[1pt]}c 
	!{\color{cyan} \VRule[1pt]}c
	!{\color{cyan} \VRule[1pt]}c 
	!{\color{cyan}\VRule[1pt]}
}\hline
	Tag & Relative & Barcode & Barcode & $v_{D}$ & $v_{tof}$ & $v_{chain}$ & Barcode & Barcode & Barcode\\
	Label & Position & Dwell & Tof & ($10^{-1}$) & ($10^{-1}$) & $\simeq\bar{v}_{scan} (10^{-1})$ & Method I & Method II & Method II\\ 
	{\color{red} Fig.~2} & {\color{red} Exact} & {\color{red} Eqn.~7} & {\color{red} Eqn.~6c} & {\color{red} Eqn.~3c} & {\color{red} Eqn.~4c} & {\color{red} Eqn.~10} & {\color{red} Dwell/Tof} & {\color{red} Dwell} & {\color{red} Tof} \\
\hline

{\color{blue} \bf{} $T_1$} & -460 & -486 $\pm$ 124 & -407 $\pm$ 45 & 2.590 $\pm$ 0.873 & 2.205 $\pm$ 0.335 & 2.374 $\pm$ 0.155 & -459 $\pm$ 17 & -461 $\pm$ 27 & -461 $\pm$ 27 \\ \hline
{\color{red} \bf{} $T_2$} & -245 & -251 $\pm$ 64 & -203 $\pm$ 25 & 2.321 $\pm$ 0.828 & 1.922 $\pm$ 0.295  & 2.374 $\pm$ 0.155 & -248 $\pm$ 16 & -249 $\pm$ 19 & -248 $\pm$ 19 \\ \hline
{\color{orange} \bf{} $T_3$}& -235 & -237 $\pm$ 60 & -191 $\pm$ 24 & 2.310 $\pm$ 0.749 & 1.874 $\pm$ 0.274 & 2.374 $\pm$ 0.155  & -236 $\pm$ 16 & -237 $\pm$ 18 & -237 $\pm$ 18 \\ \hline
{\color{magenta} \bf{} $T_4$} & -215 & -219 $\pm$ 57 & -177 $\pm$ 22 & 2.371 $\pm$ 0.829 & 1.943 $\pm$ 0.283 & 2.374 $\pm$ 0.155 & -213 $\pm$ 16 & -214 $\pm$ 17 & -214 $\pm$ 17 \\ \hline
{\color{gray} \bf{} $T_5$} & 0 & 0 & 0 & 2.509 $\pm$ 0.897 & 1.998 $\pm$ 0.276 & 2.374 $\pm$ 0.155 & 0 & 0 & 0\\ \hline
{\color{teal} \bf{} $T_6$} & 11 & 12 $\pm$ 3 & 10 $\pm$ 1 & 2.545 $\pm$ 0.870 & 1.997 $\pm$ 0.287 & 2.374 $\pm$ 0.155 & 12 $\pm$ 2 & 11 $\pm$ 2 & 10 $\pm$ 2\\ \hline
{\color{green} \bf{} $T_7$} & 82 & 93 $\pm$ 23 & 77 $\pm$ 7 & 2.626 $\pm$ 0.928 & 2.185 $\pm$ 0.353 & 2.374 $\pm$ 0.155 & 87 $\pm$ 10 & 87 $\pm$ 11 & 87 $\pm$ 11 \\ \hline
{\color{cyan} \bf{} $T_8$} & 287 & 304 $\pm$ 72 & 254 $\pm$ 26 & 2.621 $\pm$ 0.912 & 2.225 $\pm$ 0.312 & 2.374 $\pm$ 0.155 & 287 $\pm$ 17 & 287 $\pm$ 27 & 287 $\pm$ 26 \\ \hline
\end{tabular}
\end{table*}
%%%%%%%%%%%%%%%%%%%%%%%%%%%%%%%%%%%%%%%%%%%%%%%%%%%%%%%%%%%%%%
A closer look reveals 
the over/under estimation of the barcodes (columns 3 \& 4) w.r.t the theoretical value (column 1)  occurs when
$v_D$ and $v_{tof}$ (columns 5 \&  6) are greater or less than  the average velocity of the entire chain $\langle v_{chain}\rangle $ (column 7), and can be immediately discerned from Fig.~\ref{velocity}.  
Furthermore this is an {\em uncontrolled} approximation introduced in Eqn.~\ref{vmn} and depends on the contour length separation between the tags which is the unknown to be determined. We further observe that since $v_{tof}(m) < v_{chain}$ for all the tags $m=1, 8$, the barcodes are underscored. On the contrary, $v_D(m)$ for the tags are more dispersed above and below $v_{chain}$, and whenever $v_D(m) \simeq v_{chain}$, Eqn.~\ref{vmn} gives a better agreement (for ${\bf\color{orange} T_3}$ and ${\bf\color{magenta} T_4}$). If we replace the approximate velocities in Eqn.~\ref{dmn3} by the constant velocity $v_{chain}$ of the entire chain this improves the estimates significantly. This is shown in column 8 (Barcode Method-I) and discussed later. We now explain the source of discrepancy  using the non-equilibrium tension propagation theory.
\par
{\em $\bullet$~Tension Propagation (TP) Theory explains the source of discrepancy and provides a solution:}~
Unlike a rigid rod, tension propagation governs the semi-flexible chain's motion in the presence of an external bias
force. In TP theory~\cite{Sakaue_PRE_2007} and its implementation in
Brownian dynamics~\cite{CNP-Barcode-2021, Ikonen_JCP2012, Adhikari_JCP_2013} the motion of the subchain in the {\em cis} side decouples into two domains. In the vicinity of the pore, the tension front affects the motion directly while the second domain remains unperturbed, beyond the reach of the TP front. \par
\begin{figure}[ht!]
  \includegraphics[width=0.45\textwidth]{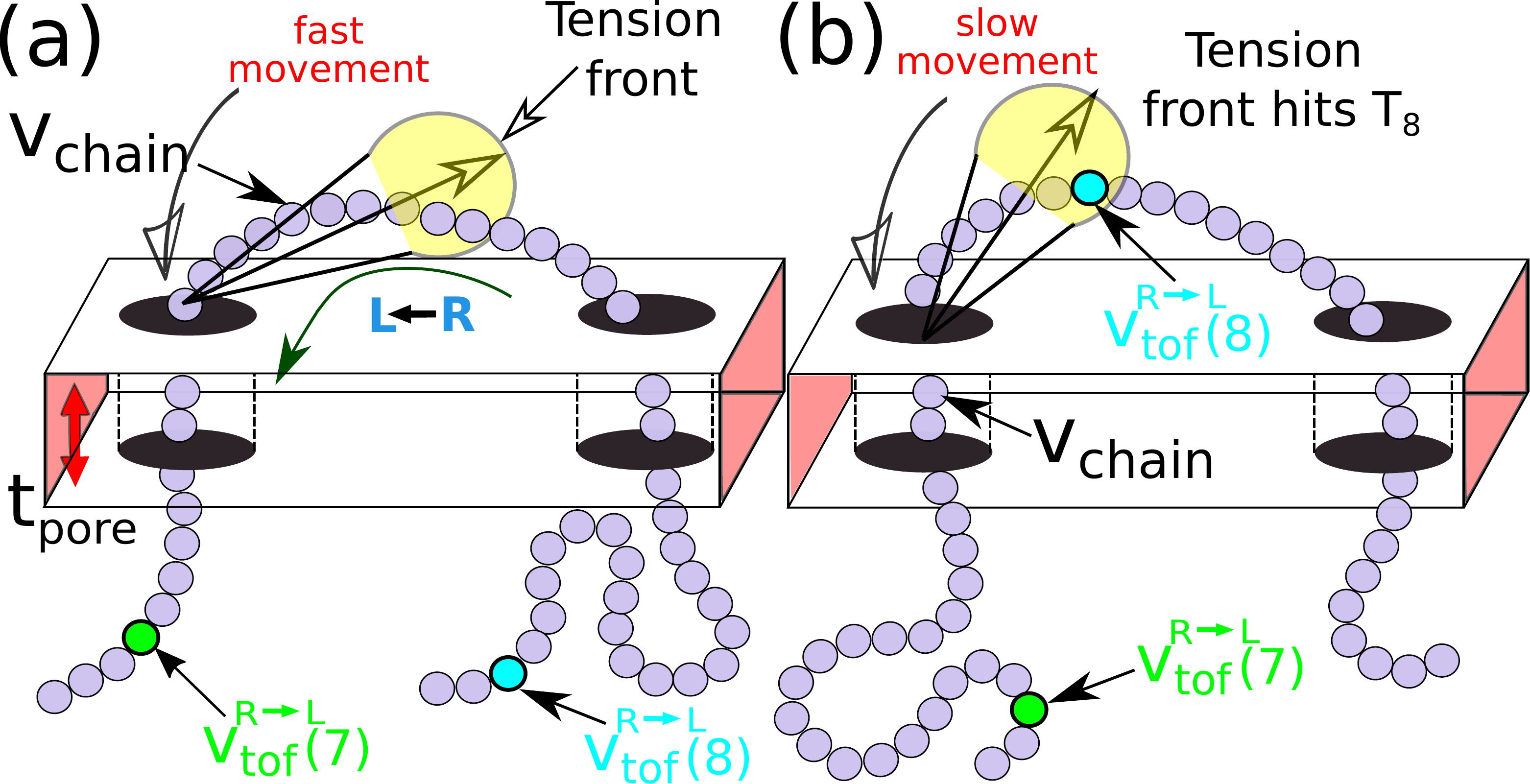}
   \caption{\small \label{tp} Demonstration of tension propagation (TP) during translocation. (a) A sudden onset of faster movement of monomers right after the translocation of {\bf \color{green} $T_7$} continues until the TP front hits {\bf \color{cyan} $T_8$}. The velocity of the monomers is comparable to the average chain velocity $v_{chain}$. (b) TP front reaches to {\bf \color{cyan} $T_8$} and slower translocation begins due to tag's inertia and larger viscous drag.}
\end{figure}

In our case, after the tag $T_m$ translocates through the pore, preceding monomers are dragged into the pore quickly by the tension front, analogous to the uncoiling effect of a rope pulled from one end. The onset of this sudden {\em faster} motion continues to grow (Fig.~\ref{tp}(a)) and reaches its maximum until the tension front hits the subsequent tag $T_{m \pm1}$ (for $R/L \rightarrow L/R$ translocation direction), having larger inertia and viscous drag. At this time (called the tension propagation time~\cite{Adhikari_JCP_2013})  the faster motion of the monomers  begins to taper down to the velocity of the tag $T_{m \pm1}$. An example of this process is shown in Fig.~\ref{tp}. This process continues from one segment to the other and %%%%%%%%%%%%%%%%%%%%%%%%%%%%%%%%%%%%%%%%%%%%%%%%%%%%%%%%%%%%
%%%%%%%%%%%%%%%%%%%%%%%%%%%%%%%%%%%%%%%%%%%%%%%%%%%%%%%%%%%%%%% 
explains oscillatory characteristics in Fig.~\ref{velocity}(c). These contour lengths of faster moving segments in between two tags are accounted for neither in  Eqn~\ref{dmn3} nor in Eqn.~\ref{d_dwell}. The experimental protocols are limited in extracting barcode information through Eqns.~\ref{dmn3} and \ref{d_dwell} (measuring current blockade time) and could be the possible source of error. \par
$\bullet$~{\em How to estimate the barcodes accurately ?~}
We now propose two methods that take into account the dynamics of the entire chain and correctly determine the barcodes and can be implemented
in a dual nanopore setup experimentally.
\par
\begin{figure}[ht!]
\centering
\includegraphics[width=0.4\textwidth]{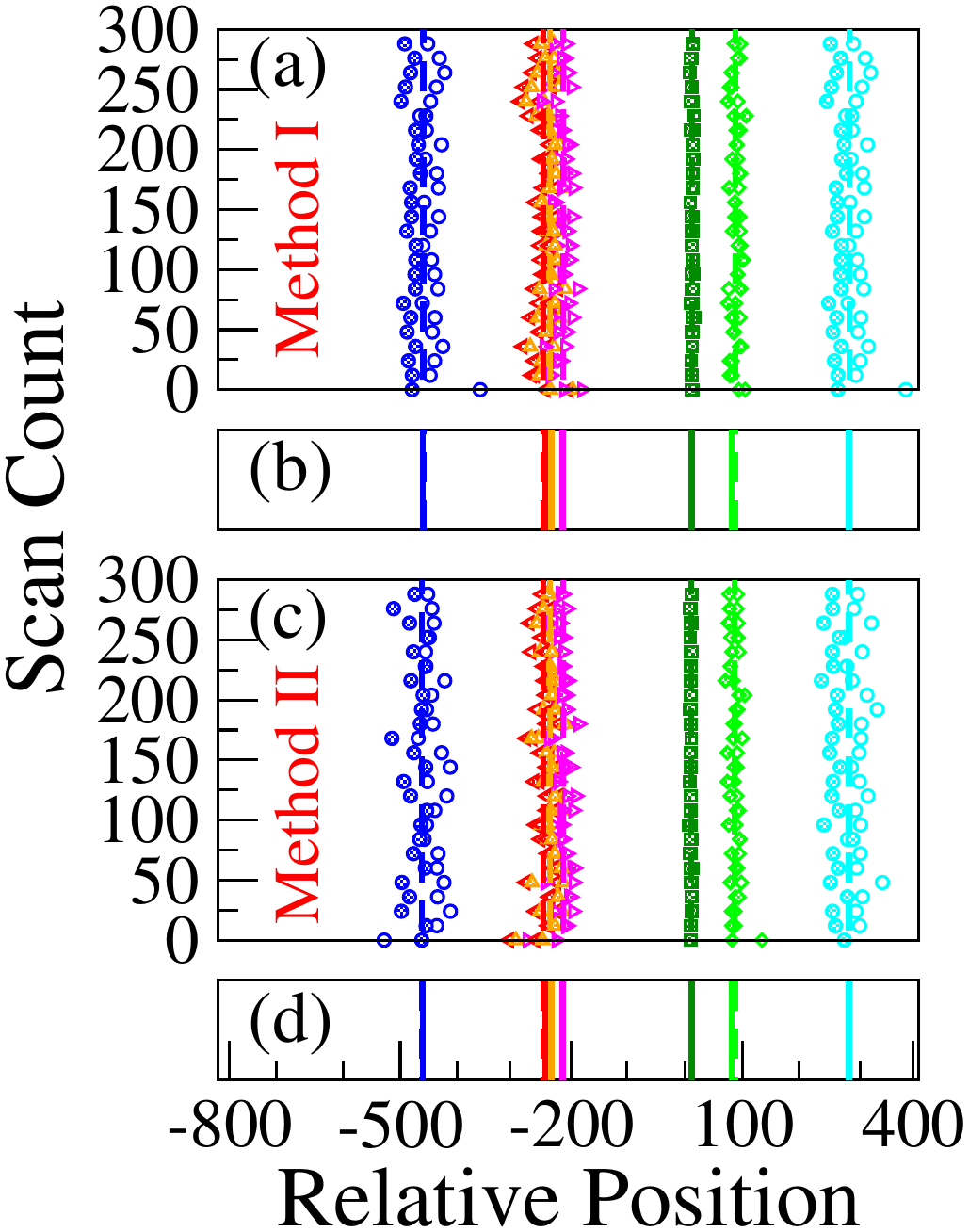}
\caption{\small \label{BC_method12} Barcodes generated using different methods using tof velocity information. The symbols have the same meanings as in Fig.~\ref{BC-dwell}.}
\end{figure}

{\em Method 1 - Barcode from known end-to-end Tag distance:}~
If the distance between the first tag {\bf \color{blue} $T_1$} 
and the last tag {\bf \color{cyan} $T_8$} $d_{18}\simeq L $, then the velocity of the segment $d_{18}$  will approximately account for the average velocity of the entire chain ($v_{chain}$) so that 
\begin{equation}
v_{chain}^{L \rightarrow R} \approx v_{18}^{L \rightarrow R} = d_{18}/(\Delta \tau)_{18}^{L \rightarrow R},
\label{v_chain}
\end{equation}
assuming we know $d_{18}$ and  $(\Delta \tau)_{18}^{L \rightarrow R}$  is the time delay of arrival at the pore between {\bf \color{blue} $T_1$} and {\bf \color{cyan} $T_8$} for ${L \rightarrow R}$ translocation. We then estimate the barcode distance  $d_{mn}^{L \rightarrow R}$ between tags $T_m$ and $T_n$ as 
\begin{equation}
  d_{mn}^{L \rightarrow R} =  v_{18}^{L \rightarrow R} \times (\Delta \tau)_{mn}^{L \rightarrow R}. 
\label{method1}
\end{equation}
The barcodes for the $ L \rightarrow R$ and $ R \rightarrow L $ translocation are shown in Fig.~\ref{BC_method12}(a). The average shown in  Fig.~\ref{BC_method12}(b) corresponds to column 8 of Table-II as mentioned earlier. This is a significant improvement compared to the usage of the average tag velocity of the chain segments. This method will work if one can put two additional tags at known distances, at or close to the two ends of the DNA being scanned. Alternately, scan time information can be used to have  a better estimate of the average velocity of the chain. In our simulation, we kept the scanning length $L_{scan}$ constant with starting and ending values ranging from $0.0976L$ to $0.902L$. By using the constant scanning length $L_{scan}$, the average scan velocity $\bar{v}_{scan}$ can be used to determine barcodes by replacing $v_{chain}\simeq v_{18}$ in Eqn.~\ref{method1} with 
\begin{equation}
\bar{v}_{\rm scan} = \frac{1}{N_{\rm scan}}\sum_{i=1}^{N_{\rm scan}} L_{\rm scan}/\tau_{\rm scan}(i),\\
\label{vscan}
\end{equation}
where $\tau_{\rm scan}(i)$ is the scan time for the $i^{th}$ event, and $N_{\rm scan}=300$.
\par
%%%%%%%%%%%%%%%%%%%%%%%%%%%%%%%%%%%%%%%%%%%%%%%%%%%%%%%%%%%%%%
\par
$\bullet${\em ~Method 2 - Barcode using two-step method:}~Having gained a better understanding of the velocities of the monomers of the dsDNA segments in between the tags we now rectify Eqns.~\ref{dmn3} and \ref{d_dwell} by taking a {\em weighted average} of the velocities of tags and DNA segment in between as follows. 
First, we estimate the approximate number of monomers $N_{mn}\simeq d_{mn}^{L \rightarrow R}/\langle b_l \rangle $
($\langle b_l \rangle$  is the bond length) by considering the tag velocities only using Eqns.~\ref{dmn3}. We further re-calculate the segment velocity accurately by incorporating weighted velocity contributions from both the tags and the monomers between the tags as follows:
%%%%%%%%%%%%%%%%%%%%%%%%%%%%%%%%%%%%%%%%%%%%%%%%%
\begin{equation}
\begin{split}
  v_{weight}^{L \rightarrow R} & = \frac{1}{N_{mn}} \Big[ n_{next} \left( v_{tof}^{L \rightarrow R}(m) + v_{tof}^{L \rightarrow R}(n) \right)  \\
  & +  \left(N_{mn}-2n_{next}\right) \bar{v}_{\rm scan} \Big].
\end{split}
\label{twostep}
\end{equation}
Here, $n_{next}$ are the number of neighboring monomers adjacent to the tags those share the same tag velocity. We checked that $n_{next} \approx 1 - 3 $ does not make a noticeable difference in the final result. 
The barcodes are finally calculated as
\begin{equation}
d_{mn}^{L \rightarrow R} = v_{weight}^{L \rightarrow R}\times (\Delta\tau)_{mn}^{L \rightarrow R}
\end{equation}
for $L \rightarrow R$ translocation and repeating the procedure for $R \rightarrow L$ translocation, shown in
Fig.~\ref{BC_method12}(c) for both ${L \rightarrow R}$ and ${R \rightarrow L}$ translocation. The average shown in Fig.~\ref{BC_method12}(d) corresponds to column 10 of Table-II. It is worth noting that (i)  in Eqn.~\ref{twostep} the tag velocities are more weighted and makes a difference when $N_{mn} $ is small, {\em i.e.}, the contour length between the tags is small, in which cases the monomers in between the tags move with almost same velocity as that of the tags. In the other limit when $N_{mn}>>1$, it is the chain velocity that dominates and one can safely ignore the velocity of the two tags (the 1st two terms in Eqn.~\ref{twostep}). Since the number of tags are only a few (8 in 1024 in our case), Eqn.~\ref{method1} works well excepting when the tags are close by. (ii) The ``two step'' weighting procedure in Eqn.~\ref{twostep} is only approximate and has room for further improvement as one can interpolate from $v_{tag}$ to $v_{chain}$ with a suitable interpolation scheme.
\par
$\bullet${\em ~Dwell time versus TOF:}~We have repeated the same protocol to correct the data from the dwell time measurement replacing $v_{tof}$ in Eqn.~\ref{twostep} by $v_{D}$ listed in column 10 of Table-II. They are practically indistinguishable excepting for short tag distances.

\par
\begin{figure}[ht!]
\includegraphics[width=0.4\textwidth]{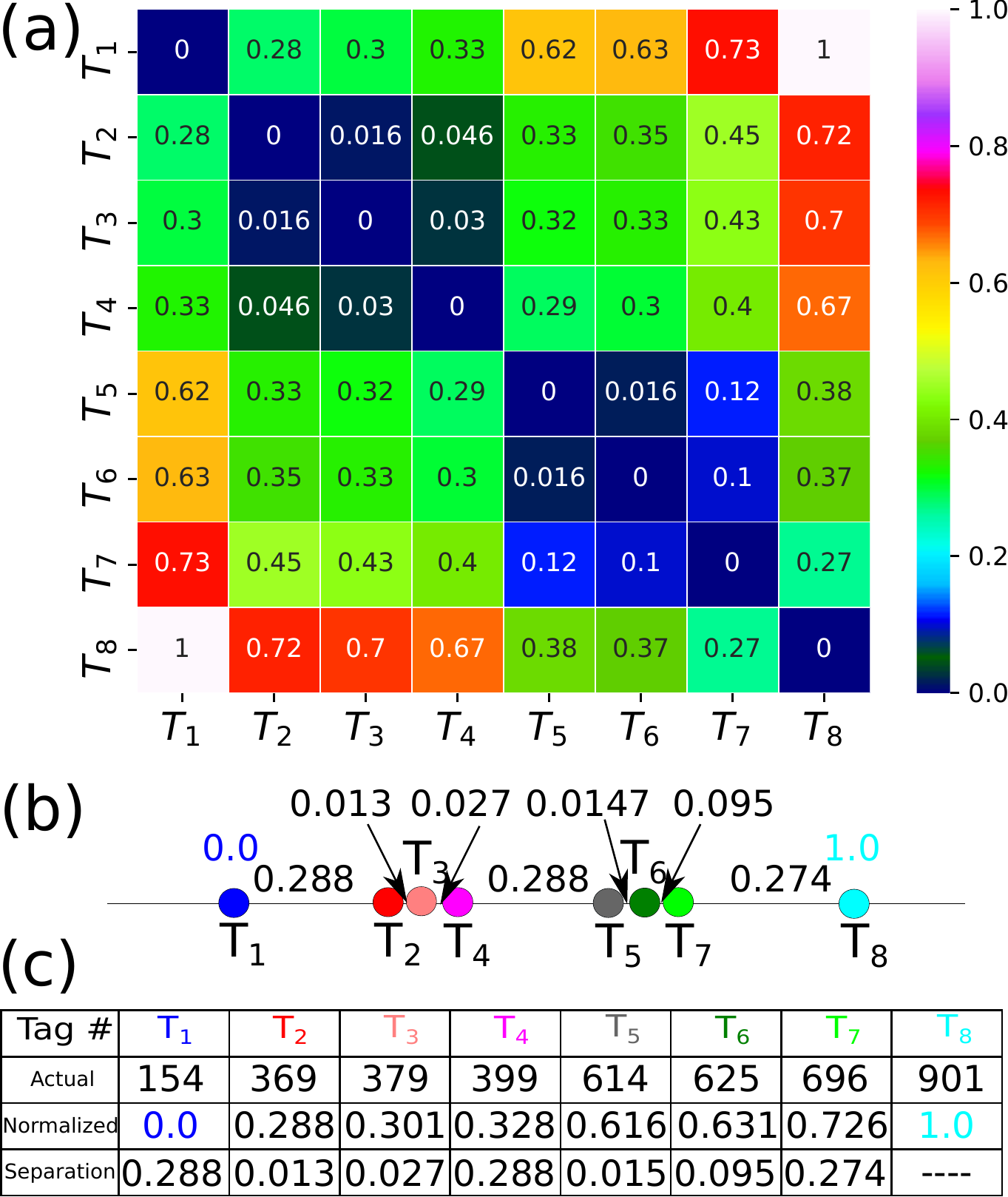}
\caption{\small \label{heatmap} (a) Heat map of the normalized tag time delay $(\tilde{\Delta}\tau)_{mn}=(\Delta\tau)_{mn} / (\Delta\tau)_{18} $ for $m=1,8$ and $n=1,8$. (b) Inter tag spacing in the normalized unit where $d_{18} = 1.0$. The normalized tag time delays are proportional to the normalized spacing distances validate the accuracy of Method I. (c) Table contains the actual tag positions and normalized inter-spacing distances.}
\end{figure}

$\bullet${\em~Tag-time-delay matrix and the sum-rule:}~If we use $v_{chain}=\rm{constant}$ to determine the barcodes as in
Eqn.~\ref{method1}, then the average {\em tag-time-delay}
$\langle (\Delta \tau)_{mn} \rangle = \frac{1}{2} \langle (\Delta\tau)_{mn}^{L \rightarrow R} + (\Delta\tau)_{mn}^{R \rightarrow L}  \rangle $ will be proportional to the barcode distances. One can then form a heat map of the normalized tag time delay $(\tilde{\Delta}\tau)_{mn}=(\Delta\tau)_{mn} / (\Delta\tau)_{18} $ in the form of a $8\times8$ matrix as shown in Fig.~\ref{heatmap}. The values in each square when multiplied by the appropriate scale factor ($\Delta\tau)_{18}\times v_{chain}$) will reproduce the barcode distances and can serve as a nice visual about the relative distances between the tags. We find that indeed it reproduces barcodes of Fig.~\ref{tags} as listed in Table-I.
%%%%%%%%%%%%%%%%%%%%%%%%%%%%%%%%%%%%%%%%%%%%%%%%%%%%%%%%%%%%%%%
%%%%%%%%%%%%%%%%%%%%%%%%%%%%%%%%%%%%%%%%%%%%%%%%%%%%%%%%%%%%%%%
We also find a ``sum-rule'' $(\tilde{\Delta}\tau)_{mn}=(\tilde{\Delta}\tau)_{mp}+(\tilde{\Delta}\tau)_{pn}$ to be satisfied. As an example, let's choose $m=1$, $n=3$ so that
$(\tilde{\Delta}\tau)_{13}= 0.30 =(\tilde{\Delta}\tau)_{12} + (\tilde{\Delta}\tau)_{23} = 0.280 + 0.016 = 0.296 \simeq 0.30$, the normalized distance between the tag $T_1$ and $T_3$. In general, one can check that
\begin{equation}
  \label{sum-rule}
(\tilde{\Delta}\tau)_{mn}= (\tilde{\Delta}\tau)_{m (m+1)}+(\tilde{\Delta}\tau)_{(m+1) (m+2)}\cdot\cdot (\tilde{\Delta}\tau)_{(n-1) n}.
\end{equation}
Thus this {\em sum-rule} can be used to measure the distance between barcodes in many different ways, reduce the uncertainties, and possibly infer information about a missing tag from the self-consistency checks using Eqn.~\ref{sum-rule}.
\par
{\em $\bullet$~Realistic pores and biases:~}Two pores in an experimental dual nanopores system are not exactly identical, about 5-10\% differences in pore diameters are reported~\cite{Reisner-Small-2020}.
Likewise, when the differential bias is reversed $\Delta \vec{f}_{LR} = \vec{f}_L-\vec{f}_R \ne \Delta \vec{f}_{RL} = \vec{f}_R-\vec{f}_L$ for translocation directions $L \rightarrow R$ and $R \rightarrow L$ respectively. We ran additional simulations by offsetting the ideal conditions and checked that the same methods (Method I and II) work.  
\par 
{\em $\bullet$~Summary and concluding remarks:}~Dual nanopore platform has immense promise and advantages compared to its single nanopore counterpart. In this Brownian dynamics simulation study, we mimicked an experimental platform and explained why extracting information from the tags only need to be corrected by taking into account the motion of the entire chain. We invoked tension propagation theory to explain the velocity distribution of the entire chain as a function of the monomer index. The protein tags introduce oscillation on the uniform velocity of the chain that depends on the tension propagation time from one tag to the other. We have checked that the information obtained from the time of flight data is more accurate compared to the dwell time data from the individual pores. We further discovered that the most reliable quantity is the {\em tag-time-delay} of the successive barcodes to arrive at the L/R pore. When the distance between the tags is large the {\em tag-time-delay} will straight translate to genomic length excepting for those cases when the tags are close by. Our two-step interpolation scheme will overcome this issue. This is due to the fact that roughly it is the average velocity of the entire chain and not the average velocity of the segment between two tags need to be used to calculate the barcode distances. The heat-map of the normalized {\em tag-time-delay} provides and the corresponding {\em sum-rule} are the direct proof of the efficacy of this method. This study also indicates how to improve the measurement protocol. With some prior information about the tags if one can selectively attach heavier molecules at the tag positions so that the velocities will produce sharp dips on the velocity profile of the entire chain, then the procedure will be more accurate. We believe that this study will be immensely useful for designing future double nanopore platforms so that the data in the time domain can be translated to unravel fine structures of genomic lengths.
%%%%%%%%%%%%%%%%%%%%%%%%%%%%%%%%%%%%%%%%%%%%%%%%%%%%%%%%%%%%%%%
\vskip 0.2truecm
\par
{\em $\bullet$~Notes:}~
The authors declare no competing financial interest.
%%%%%%%%%%%%%%%%%%%%%%%%%%%%%%%%%%%%%%%%%%%%%%%%%%%%%%%%%%%%%%%
\vskip 0.2truecm
\par
{\em $\bullet$~Acknowledgments:}~The research at UCF has been supported by the grant number 1R21HG011236-01 from the National Human Genome Research Institute at the National Institute of Health. We thank Walter Reisner and An Vong for discussions about double nanopore experiments. All computations were carried out at the UCF's high performance computing platform STOKES.

\vskip 1.72truecm
%%%%%%%%%%%%%%%%%%%%%%%%%%%%%%%%%%%%%%%%%%%%%%%%%%%%%%%%%%%%%%%


\begin{thebibliography}{50}

\bibitem{barcode_CoxI} Hebert, P. D. N., Ratnasingham, S. $\&$ de Waard, J. R. Barcoding Animal Life: Cytochrome c Oxidase Subunit 1 Divergences among Closely Related Species. Proc. R. Soc. Lond. B {\bf 270}, (2003).

\bibitem{barcode_Hebert} Hebert, P. D. N., Cywinska, A., Ball, S. L. $\&$ deWaard, J. R. Biological Identifications through DNA Barcodes.  Proc. R. Soc. Lond. B  {\bf 270} 313-321 (2003).

\bibitem{barcode_cryptic_species} Hebert, P. D. N., Penton, E. H., Burns, J. M., Janzen D. H. $\&$ Hallwachs, W. Ten Species in One: DNA Barcoding Reveals Cryptic Species in the Neotropical Skipper Butterfly Astraptes Fulgerator. Proceedings of the National Academy of Sciences {\bf 101}, 14812-14817 (2004). 

\bibitem{barcode_taxonomy} Schindel, D. E. $\&$ Miller, S. E. DNA Barcoding a Useful Tool for Taxonomists. Nature {\bf 435}, 17-17 (2005).
  
\bibitem{barcode_seafood} Wong, E. H.-K. $\&$ Hanner, R. H. DNA Barcoding Detects Market Substitution in North American Seafood. Food Research International {\bf 41}, 828-837 (2008).

\bibitem{barcode_bio_diversity} Vernooy, R., Haribabu, E., Muller, M. R., Vogel, J. H., Hebert, P. D. N., Schindel, D. E., Shimura, J. $\&$ Singer, G. A. C. Barcoding Life to Conserve Biological Diversity: Beyond the Taxonomic Imperative. PLoS Biol {\bf 8}, e1000417 (2010).  

\bibitem{barcode_MiniON} Chang, J. J. M., Ip, Y. C. A., Bauman, A. G. $\&$ Huang, D. MinION-in-ARMS: Nanopore Sequencing to Expedite Barcoding of Specimen-Rich Macrofaunal Samples From Autonomous Reef Monitoring Structures. Front. Mar. Sci. {\bf 7}, (2020).

\bibitem{Dekker-2016} Pud, S., Chao, S.-H., Belkin, M., Verschueren, D., Huijben, T., van Engelenburg, C., Dekker, C. $\&$ Aksimentiev, A. Mechanical Trapping of DNA in a Double-Nanopore System. Nano Lett. {\bf 16}, 8021-8028 (2016).

\bibitem{Reisner-Small-2018} Zhang, Y., Liu, X., Zhao, Y., Yu, J.-K., Reisner, W. $\&$ Dunbar, W. B. Single Molecule DNA Resensing Using a Two-Pore Device. Small {\bf 14}, 1801890 (2018). 
  
\bibitem{Reisner-Small-2019}
Liu, X., Zhang, Y., Nagel, R., Reisner, W. $\&$ Dunbar, W. B. Controlling DNA Tug-of-War in a Dual Nanopore Device. Small {\bf 15}, 1901704 (2019).

\bibitem{Reisner-Small-2020} Liu, X., Zimny, P., Zhang, Y., Rana, A., Nagel, R., Reisner, W. $\&$ Dunbar, W. B. Flossing DNA in a Dual Nanopore Device. Small {\bf 16}, 1905379 (2020).
  
\bibitem{Bhattacharya_Seth_2020} Bhattacharya, A. $\&$ Seth, S. Tug of War in a Double-Nanopore System. Phys. Rev. E {\bf 101}, (2020).

\bibitem{Seth-JCP-2020} Seth, S. $\&$ Bhattacharya, A. Polymer Escape through a Three Dimensional Double-Nanopore System. The Journal of Chemical Physics {\bf 153}, 104901 (2020).

\bibitem{Aksimentiev-2020}
Choudhary, A., Joshi, H., Chou, H.-Y., Sarthak, K., Wilson, J., Maffeo, C. $\&$ Aksimentiev, A. High-Fidelity Capture, Threading, and Infinite-Depth Sequencing of Single DNA Molecules with a Double-Nanopore System. ACS Nano {\bf 14}, 15566-15576 (2020).

\bibitem{Review1}
Muthukumar, M. Polymer Translocation. (2016) doi:10.1201/b10901.

\bibitem{Review2}
Milchev, A. Single-polymer dynamics under constraints: scaling theory and computer experiment. J. Phys.: Condens. Matter {\bf 23}, 103101 (2011).
  
\bibitem{Review3}
Palyulin, V. V., Ala-Nissila, T. $\&$ Metzler, R. Polymer translocation: the first two decades and the recent diversification. Soft Matter {\bf 10}, 9016-9037 (2014).

\bibitem{Sakaue_PRE_2007} Sakaue, T. Nonequilibrium Dynamics of Polymer Translocation and Straightening. Phys. Rev. E {\bf 76}, (2007).

\bibitem{Ikonen_JCP2012}
Ikonen, T., Bhattacharya, A., Ala-Nissila, T. $\&$ Sung, W. Influence of Non-Universal Effects on Dynamical Scaling in Driven Polymer Translocation. The Journal of Chemical Physics {\bf 137}, 085101 (2012). 

\bibitem{Adhikari_JCP_2013} Adhikari, R. $\&$ Bhattacharya, A. Driven Translocation of a Semi-Flexible Chain through a Nanopore: A Brownian Dynamics Simulation Study in Two Dimensions. The Journal of Chemical Physics {\bf 138}, 204909 (2013).

\bibitem{CNP-Barcode-2021} Seth, S. $\&$ Bhattacharya, A. DNA Barcodes using a Cylindrical Nanopore. Preprint at \url{https://arxiv.org/abs/2102.03464} (2021).
\end{thebibliography}
\end{document}